\documentclass{jps-cp}
\usepackage{txfonts} 
\usepackage{bm}
\usepackage{bbm}

\def\gtap{\ \raise.3ex\hbox{$>$\kern-.75em\lower1ex\hbox{$\sim$}}\ }
\def\ltap{\ \raise.3ex\hbox{$<$\kern-.75em\lower1ex\hbox{$\sim$}}\ }

\title{$\eta$ photoproduction off the deuteron and low-energy $\eta$-nucleon interaction}

\author{S.X. \textsc{Nakamura}$^{1}$, H. \textsc{Kamano}$^{2}$, and T. \textsc{Ishikawa}$^{3}$}

\inst{$^{1}$Laborat\'orio de F\'isica Te\'orica e Computacional - LFTC, 
Universidade Cruzeiro do Sul, S\~ao Paulo, SP 01506-000, Brazil\\
$^{2}$Research Center for Nuclear Physics, Osaka University, Ibaraki, Osaka
567-0047, Japan\\
$^{3}$Research Center for Electron Photon Science (ELPH), 
Tohoku University, Sendai, Miyagi 982-0826, Japan}

\email{sxnakamura@gmail.com}

\recdate{December 25, 2018}

\abst{
We study $\eta$ photoproduction off the deuteron ($\gamma d\to\eta pn$)
at a special kinematics:
$\sim 0.94$~GeV of the photon beam energy 
and $\sim 0^\circ$ of the scattering angle of the proton.
This kinematics is ideal to extract
the low-energy $\eta$-nucleon scattering parameters such as 
$a_{\eta N}$ (scattering length) and 
$r_{\eta N}$ (effective range)
because the $\eta$-nucleon elastic scattering is significantly enhanced.
We show that if a ratio $R$, the $\gamma d\to\eta pn$ cross section
divided by the $\gamma p\to\eta p$ cross section convoluted with the
proton momentum distribution in the deuteron, 
is measured with 5\% error, 
${\rm Re}[a_{\eta N}]$ (${\rm Re}[r_{\eta N}]$) 
can be determined at the precision of $\sim\pm$0.1~fm ($\sim\pm$0.5~fm),
significantly narrowing down the currently estimated range of the
parameters.
The measurement is ongoing at the Research Center
for Electron Photon Science (ELPH), Tohoku University.
}

\kword{meson photoproduction, hadron-hadron interaction, $\eta$-mesic
nuclei}

\begin{document}

\begin{flushright}
\vspace*{-30mm}
\hspace*{-50mm}LFTC-18-16/37
\end{flushright}
\vspace*{20mm}

\maketitle

\section{Introduction}

The low-energy $\eta$-nucleon interaction
can be characterized with two parameters,
the scattering length $a_{\eta N}$ and effective range $r_{\eta N}$.
The existence of exotic $\eta$-mesic nuclei largely depends on
$a_{\eta N}$ that determines the attractive or repulsive nature of 
the low-energy $\eta N$ interaction~\cite{etan9}.
However, $a_{\eta N}$ has not been well determined yet.
Previous works have attempted to extract $a_{\eta N}$ and $r_{\eta N}$ 
by analyzing the $\pi N\to \pi N, \eta N$ and $\gamma N\to \pi N, \eta N$
reaction data~\cite{etan9}, 
and also the $pn \to \eta d$ reaction data~\cite{etan5}.
These analyses gave fairly consistent results for 
the imaginary parts of $a_{\eta N}$ and $r_{\eta N}$ which are within
${\rm Im} [a_{\eta N}]=0.2$--0.3~fm 
and 
${\rm Im} [r_{\eta N}]=-1$--0~fm, 
respectively~\cite{etan9}.
However, their real parts are not well-determined:
${\rm Re} [a_{\eta N}]=0.2$--0.9~fm and
${\rm Re} [r_{\eta N}]=-6$ to +1~fm.
The large model-dependence in the real parts
stems from the difficulty of isolating the $\eta N$ scattering amplitudes 
from other mechanisms involved in the reactions analyzed.

\begin{figure}[t]
\includegraphics[clip,width=1\textwidth]{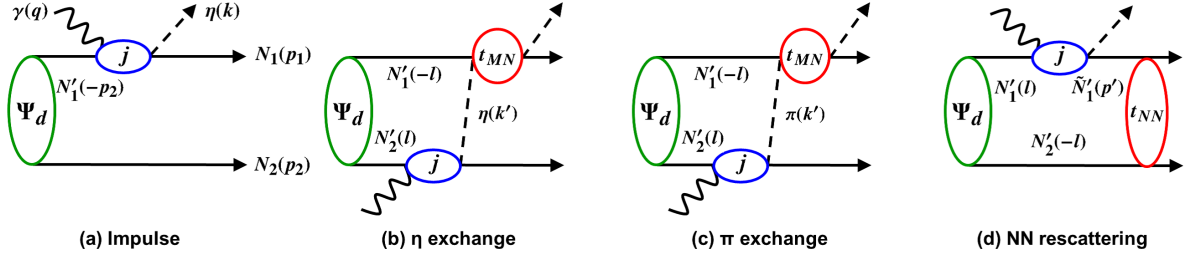}
\caption{\label{fig:diag}
Reaction mechanisms for $\gamma d\to\eta N_1 N_2$:
(a) impulse, (b) $\eta$-exchange, (c) $\pi$-exchange, and
(d) $NN$ rescattering mechanisms.
The figure taken from Ref.~\cite{ref0}. Copyright (2017) APS
}
\end{figure}

An ongoing 
$\eta$ photoproduction experiment~\cite{plan}
at the Research Center
for Electron Photon Science (ELPH), Tohoku University
is designed to overcome the difficulty of determining $a_{\eta N}$
by utilizing a special kinematics.
In this experiment, 
a photon beam with $E_\gamma\sim 0.94$~GeV hits 
a deuteron target
and the recoil proton from $\gamma d\to \eta pn$
is detected at 
$\theta_p\sim 0^\circ$.
At this kinematics, 
an $\eta$ produced from a quasi-free proton 
is almost at rest, and thus it would
interact strongly with the spectator neutron.
On the other hand,
the struck proton goes away with a large momentum, and thus
it would not interact with the $\eta$ and neutron.
This seems an ideal kinematical condition, 
referred to as the ELPH kinematics,
to determine the low-energy $\eta N$ scattering parameters.
The present theoretical analysis~\cite{ref0} will show that
a combined cross-section data for $\gamma d\to \eta pn$ and $\gamma p\to \eta p$
expected to be taken in the ELPH experiment
would indeed lead to significant reduction of the current uncertainty 
of $a_{\eta N}$ and $r_{\eta N}$.

\section{Model}

We study $\gamma d \to \eta p n$ relevant to the ELPH experiment
with a model based on the impulse and the first-order rescattering
mechanisms as illustrated in Fig.~\ref{fig:diag}.
The $\eta$-exchange mechanism [Fig.~\ref{fig:diag}(b)]
contains the $\eta N \to \eta N$ subprocess we are interested in,
while the other mechanisms (the impulse [Fig.~\ref{fig:diag}(a)], 
$\pi$-exchange [Fig.~\ref{fig:diag}(c)],
and $NN$-rescattering [Fig.~\ref{fig:diag}(d)] mechanisms)
are background processes.
The model must be built with reliable amplitudes 
for elementary $\gamma N \to MN$, $MN\to M'N$, and $NN\to NN$ processes
with $M^{(\prime)}$=$\pi,\eta$, 
as well as with a realistic deuteron wave function.
By doing so, 
we can reliably isolate the amplitude for the $\eta N \to \eta N$ subprocess
from data using 
well-predicted background contributions.
Regarding $\gamma N \to MN$ and $MN \to M'N$ amplitudes,
we employ those generated with a dynamical coupled-channels (DCC)
model~\cite{knls13,knls16}.
The DCC model is a multichannel unitary model for the $\pi N$ and $\gamma N$
reactions in the nucleon resonance region. 
It was constructed fitting $\sim 27,000$ data points, and 
successfully describes~\cite{knls13,knls16}
$\pi N \to \pi N, \pi\pi N, \eta N, K\Lambda, K\Sigma$ and
$\gamma N \to \pi N, \pi\pi N, \eta N, K\Lambda, K\Sigma$ reactions
over the energy region from the thresholds up to 
$\sqrt{s}\lesssim 2.1$~GeV.
For example, the DCC model describes
$\gamma p\to \eta p$ differential cross sections 
in a very good agreement with
data over the energy
region relevant to the following calculations of $\gamma d \to \eta p n$.
This confirms that
the most important $\gamma p \to \eta p$ amplitudes among the
elementary amplitudes for describing $\gamma d \to \eta p n$
have been well tested by the data.
This DCC model predicts the $\eta N$
scattering parameters
to be $a_{\eta N} = 0.75 + 0.26i$~fm and $r_{\eta N} = -1.6 - 0.6i$~fm,
which are consistent with the previously estimated ranges.
As for the deuteron wave function and the $NN$ scattering amplitudes, 
we use the CD-Bonn potential~\cite{cdbonn} to generate them.

\section{Result}
\begin{figure}[t]
 \includegraphics[clip,width=0.49\textwidth]{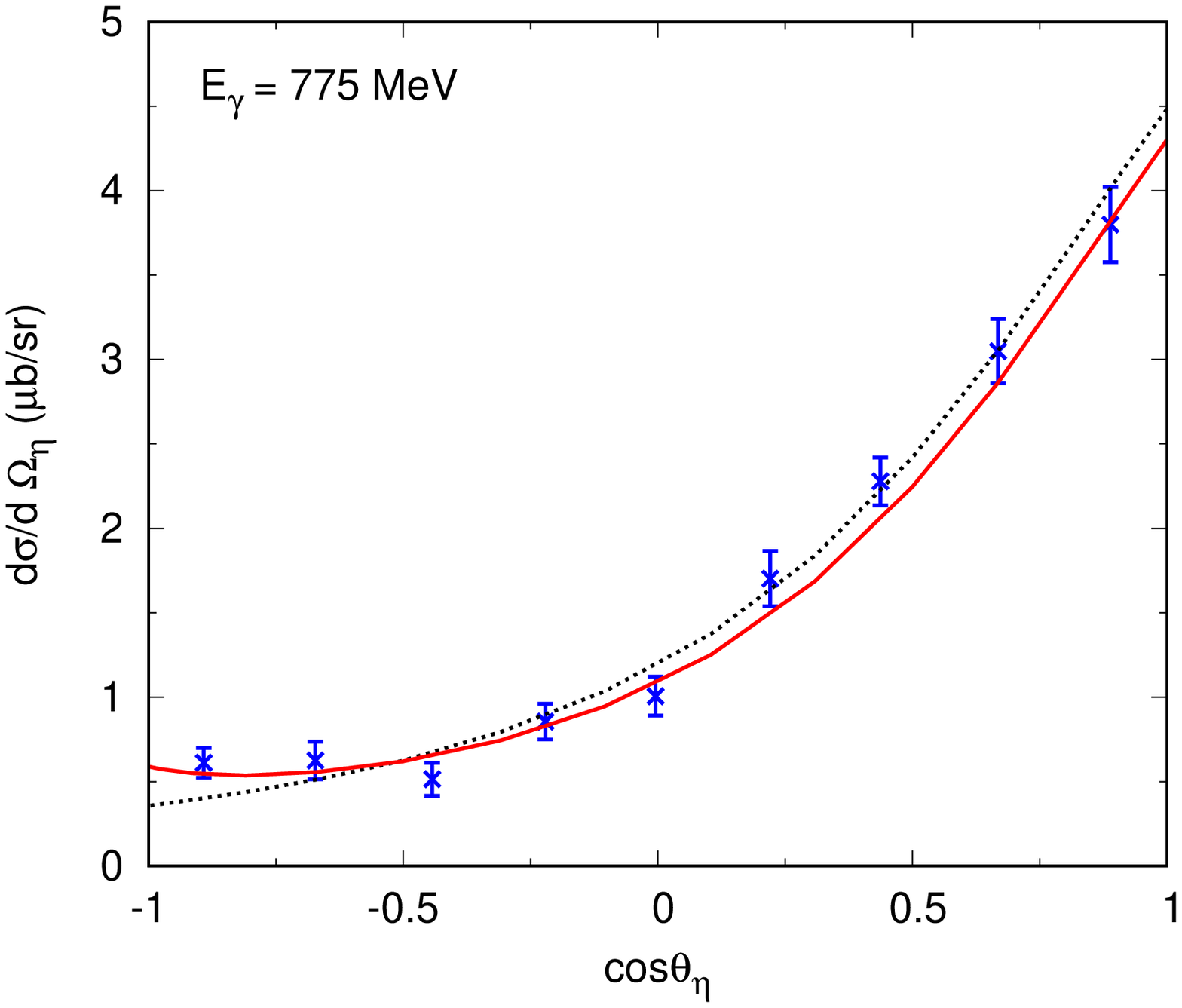}
 \includegraphics[clip,width=0.49\textwidth]{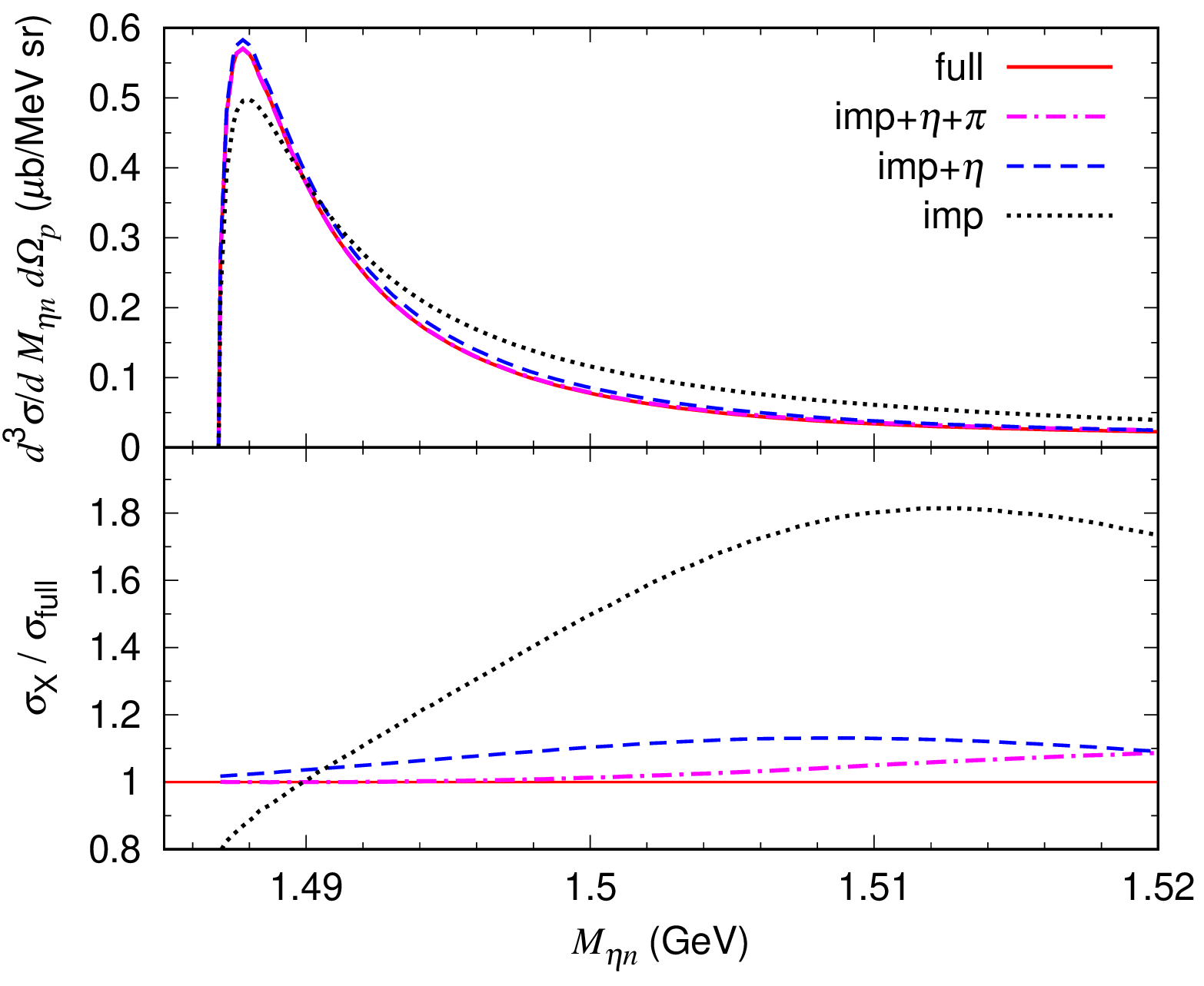}
\caption{\label{fig:eta-data}
[Left]
Predicted differential cross sections for $\gamma d\to \eta pn$.
The dotted curve is from the impulse approximation, and
the solid curve includes rescattering mechanisms in addition.
The data are from Ref.~\cite{eta-data}.
[Right-Top]
Differential cross section for $\gamma d \to \eta pn$
at $E_\gamma = 0.94$~GeV and $\theta_p=0^\circ$.
The impulse approximation (dotted curve), 
the impulse and $\eta$-exchange mechanisms (dashed curve),
the impulse, $\eta$- and $\pi$-exchange mechanisms (dash-dotted curve),
and the full calculation (solid curve).
[Right-Bottom] Ratios of the differential cross sections from the various
sets of the mechanisms to those from the full calculation.
The figures taken from Ref.~\cite{ref0}. Copyright (2017) APS
}
\end{figure}

We can make a parameter-free prediction for
the $\gamma d\to\eta pn$ cross sections using the model described above.
We can thus assess the validity of the model by
confronting our model predictions
with existing data.
In Fig.~\ref{fig:eta-data}(left), we show 
$d\sigma/d\Omega_\eta$
at $E_\gamma = 775$~MeV
from our DCC-based model 
with and without the rescattering contributions along with the data.
Our parameter-free 
prediction is found to be in an excellent agreement with the data.
The $\eta N\to \eta N$ rescattering gives
a slight enhancement in the backward direction, 
which is important for this nice agreement. 
A similar DCC-based model for $\gamma d\to\pi NN$~\cite{gd-piNN1,gd-piNN2}
also gives predictions that agree well with data by taking account of
significant rescattering effects.

\begin{figure}[t]
\begin{minipage}[t]{75mm}
\includegraphics[clip,width=1\textwidth]{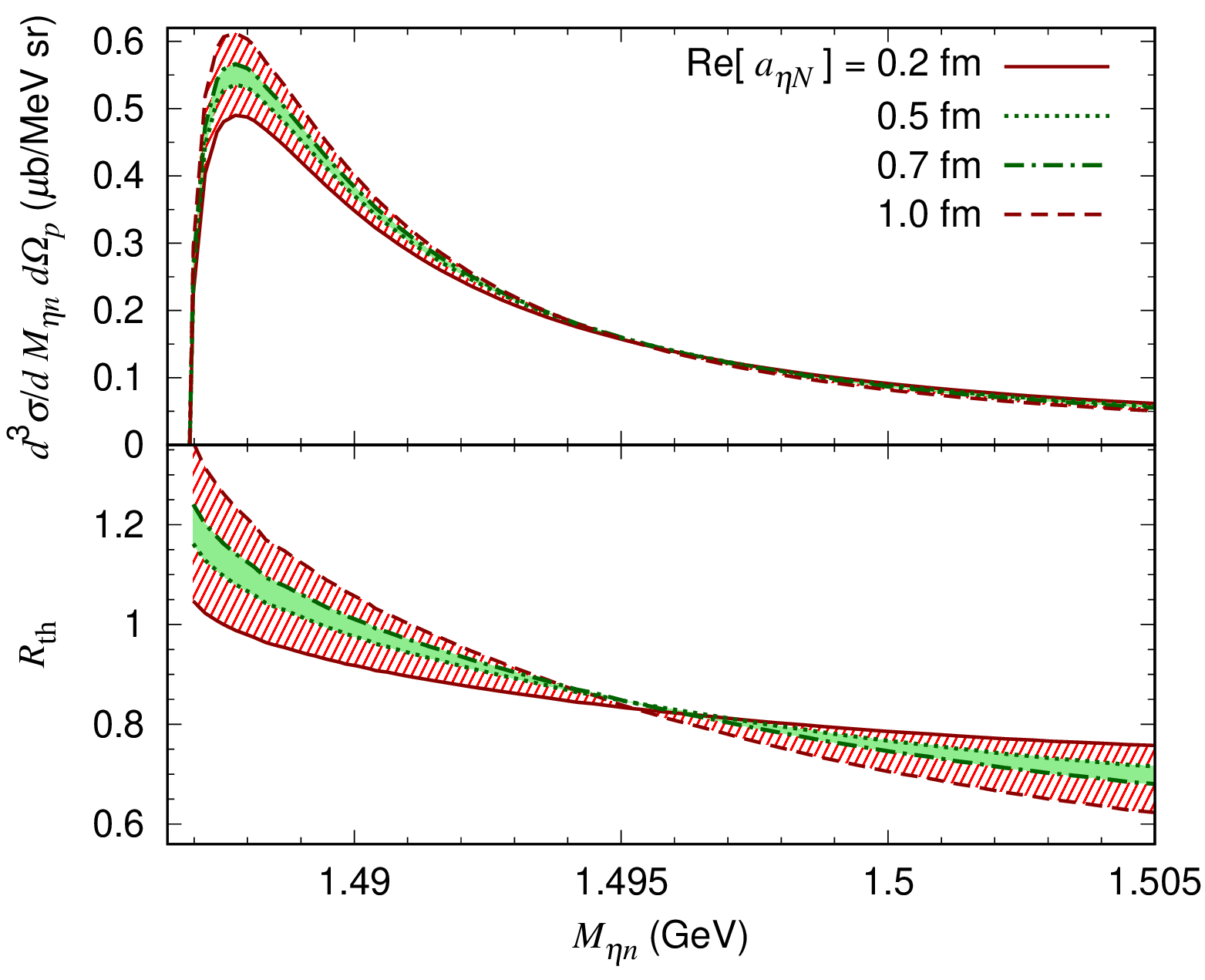}
\caption{
(Top) Re[$a_{\eta n}$]-dependence of $\gamma d\to \eta pn$ differential
 cross sections from the full model;
$E_\gamma=0.94$~GeV and $\theta_p=0^\circ$.
The curves correspond to
${\rm Re}[a_{\eta n}]=0.2$, 0.5, 0.7, and 1.0 fm;
${\rm Im}[a_{\eta n}]=0.25$~fm and $r_{\eta n}=0$.
(Bottom) The ratio $R_{\rm th}$ given by Eq.~(\ref{eq:Ratio})
calculated with different values of Re[$a_{\eta n}$].
Figure taken from Ref.~\cite{ref0}. Copyright (2017) APS
}
\label{fig:eta-a} 
\end{minipage}
\hspace{3mm}
\begin{minipage}[t]{75mm}
\includegraphics[clip,width=1\textwidth]{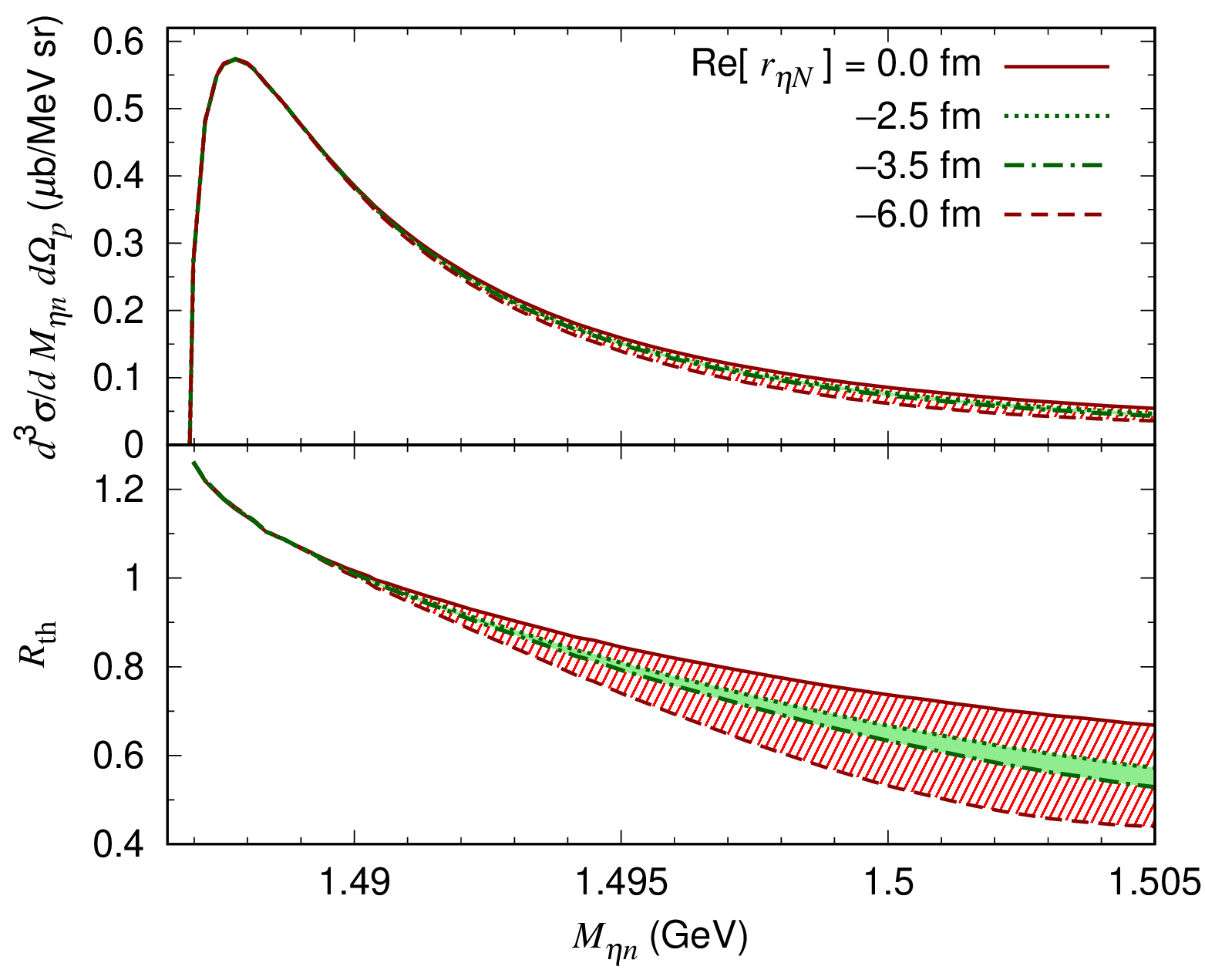}
\caption{\label{fig:eta-r} 
Presentation similar to Fig.~\ref{fig:eta-a}, but obtained with
Re$[r_{\eta n}]=0$~fm (solid), $-$2.5~fm (dotted),  $-$3.5~fm (dash-dotted), 
and $-$6~fm (dashed); 
$a_{\eta n}=0.75+0.26i$~fm
and ${\rm Im}[r_{\eta n}]=0$~fm.
The figure taken from Ref.~\cite{ref0}. Copyright (2017) APS
}
\end{minipage}
\end{figure}

We now move on to
the $\gamma d\to\eta pn$ reaction at the ELPH kinematics
($E_\gamma=0.94$~GeV and $\theta_p=0^\circ$).
In Fig.~\ref{fig:eta-data}(right-top),
the predicted threefold differential cross section, $d^3\sigma/dM_{\eta n} d\Omega_p$,
are presented as a function of the $\eta$-neutron invariant mass $M_{\eta n}$.
The impulse mechanism [Fig.~\ref{fig:diag}(a)] 
including the $\gamma p\to \eta p$ ($\gamma n\to \eta n$) amplitudes
gives the dominant (negligible) contribution.
A substantial contribution is from 
the $\eta$-exchange mechanism [Fig.~\ref{fig:diag}(b)],
and the cross sections including the impulse mechanisms only
are changed by $-$40 to +20\% 
[difference between the dashed and dotted curves in Fig.~\ref{fig:eta-data}(right-bottom)].
The $\pi$-exchange [Fig.~\ref{fig:diag}(c)] 
contribution is smaller, suppressing the cross sections
by $\lesssim$ 9\%
(difference between the dashed and dash-dotted curves).
The $NN$ rescattering [Fig.~\ref{fig:diag}(d)] contribution
(deviation of the dash-dotted curve from 1)
is very small for $M_{\eta n}\lesssim 1.5$~GeV.
This feature is what we expect to find in this special kinematics, 
and indicates that the proton essentially does not interact with
the $\eta n$ system. Thus
multiple rescatterings beyond the first-order rescattering 
[Figs.~\ref{fig:diag}(b)--\ref{fig:diag}(d)]
should be safely neglected for $M_{\eta n}\lesssim 1.5$~GeV.
We have also examined
an off-shell momentum effect associated with the $\eta n\to \eta n$
scattering amplitude and found it very small.
Because we are interested in a $M_{\eta n}$ region close to the
threshold, higher partial waves for 
the $\eta n\to \eta n$ amplitudes are negligible.
Therefore, we modify the full $\gamma d \to \eta p n$ model by 
replacing the $\eta n$ scattering amplitude with
the $S$-wave one parametrized with $a_{\eta N}$ and $r_{\eta N}$.
These scattering parameters are determined by
analyzing the forthcoming ELPH data.

The ELPH data will be given in a form of the ratio,
denoted by $R_{\rm expt}$,
of the measured cross sections for $\gamma d\to \eta pn$ divided by those
for $\gamma p\to \eta p$ convoluted with the proton momentum distribution
in the deuteron.
This is for removing systematic uncertainties of the acceptance from the detector coverage.
Therefore, from the theoretical side,
we need to calculate the corresponding quantity given by 
\begin{eqnarray}
\label{eq:Ratio}
R_{\rm th}(M_{\eta n}) = {d^3\sigma_{\rm full}/dM_{\eta n} d\Omega_p |_{\theta_p=0^\circ}
\over d^3 \sigma_{\rm imp}/dM_{\eta n} d\Omega_p|_{\theta_p=0^\circ}} 
\ ,
\end{eqnarray}
where $\sigma_{\rm full}$ ($\sigma_{\rm imp}$) is 
calculated with
the full model (the impulse term only).
Now the question is 
how sensitive $R_{\rm th}$ is against changing
$a_{\eta N}$ and $r_{\eta N}$.
Also, we are interested in what is the required precision of 
$R_{\rm expt}$ for significantly reducing the current uncertainties of 
$a_{\eta N}$ and $r_{\eta N}$.

First ${\rm Re}[a_{\eta N}]$ is changed over
0.2 -- 1.0~fm, 
with ${\rm Im}[a_{\eta N}]=0.25$~fm and $r_{\eta N}=0$~fm being fixed.
The resulting cross sections cover the red striped region 
as shown in Fig.~\ref{fig:eta-a}(top), within
the considered ELPH kinematics and $M_{\eta n}\le 1.505$~GeV.
The ratio $R_{\rm th}$ also changes accordingly as 
shown in Fig.~\ref{fig:eta-a}(bottom);
$R_{\rm th}$ shows more clearly
the sensitivity to the variation of ${\rm Re}[a_{\eta N}]$.
The cross section and thus $R_{\rm th}$
changes by $\sim$25\% at the quasi-free (QF)
peak position of $M_{\eta n}\sim 1.488$~GeV,
as indicated by the width of the striped band.
We also show the green solid bands
that have the widths of $\sim$5\% at the QF peak. 
This green band
is covered by our model when ${\rm Re}[a_{\eta N}]$ is varied by $\pm 0.1$~fm
from 0.6~fm.
This means that
$R_{\rm expt}$ data of 5\% error per MeV bin
can determine ${\rm Re}[a_{\eta N}]$ 
at the precision of $\sim\pm 0.1$~fm, 
significantly reducing the currently estimated range.
Data of $R_{\rm expt}$ with this precision is expected to be taken in 
the ongoing ELPH experiment~\cite{plan}.

Next ${\rm Re}[r_{\eta N}]$ is varied over 
$-$6 -- 0~fm which is the currently estimated range,
while
the scattering length being fixed at
the value from the latest DCC analysis~\cite{knls16},
$a_{\eta n}=0.75+0.26i$~fm;
${\rm Im} [r_{\eta N}]=0$~fm.
Accordingly, the cross section and $R_{\rm th}$ change over 
the red striped region in Fig.~\ref{fig:eta-r}.
The effect of changing $r_{\eta N}$ is visible 
at $\sim$5~MeV above the $\eta N$ threshold.
The ratio $R_{\rm th}$ at $M_{\eta n}=1.5$~GeV 
changes by $\sim$30\% ($\sim$5\%)
when ${\rm Re}[r_{\eta N}]$ is changed over
$-$6 -- 0~fm ($-$3.5 to $-$2.5~fm)
as indicated by the red striped (green solid) band.
Therefore, 
${\rm Re}[r_{\eta N}]$ at the precision of $\lesssim\pm$0.5~fm, 
which is significantly improved precision over the current estimates,
can be obtained by measuring 
$R_{\rm expt}$ data of 5\% error per MeV bin.
%

%

%
%
%
%
%

\begin{center}
{\large\bf Acknowledgments}
\end{center}
This work is in part supported by 
Funda\c{c}\~ao de Amparo \`a Pesquisa do Estado de S\~ao Paulo (FAPESP),
Process No.~2016/15618-8,
and by JSPS KAKENHI Grant Number JP18K03632.

\end{document}